\documentclass{ifacconf}

\usepackage{graphicx}      
\usepackage{amsmath}
\usepackage{amssymb}
\usepackage{psfrag}
\usepackage{epstopdf}
\usepackage{booktabs}
\usepackage{natbib} 
\usepackage{algorithmic}

\begin{document}
\begin{frontmatter}

\title{A Human-Computer Interface Design for Quantitative Measure of Regret Theory} 


\author{Longsheng~Jiang, Yue~Wang} 
\thanks{Research supported in part by
the National Science Foundation under Grant No. CMMI-1454139}

\address{Clemson University
   Clemson, SC 29632 USA (e-mail: longshj, yue6@clemson.edu).}

\begin{abstract}                
Regret theory is a theory that describes human decision-making under risk. The key of obtaining a quantitative model of regret theory is to measure the preference in humans' mind when they choose among a set of options. Unlike physical quantities, measuring psychological preference is not procedure invariant, i.e. the readings alter when the methods change. In this work, we alleviate this influence by choosing the procedure compatible with the way that an individual makes a choice. We believe the resulting model is closer to the nature of human decision-making. The preference elicitation process is decomposed into a series of short surveys to reduce cognitive workload and increase response accuracy. To make the questions natural and familiar to the subjects, we follow the insight that humans generate, quantify and communicate preference in natural language. The fuzzy sets theory is hence utilized to model responses from subjects. Based on these ideas, a graphical human-computer interface (HCI) is designed to articulate the information as well as to efficiently collect human responses. The design also accounts for human heuristics and biases, e.g. range effect and anchoring effect, to enhance its reliability.  The overall performance of the survey is satisfactory because the measured model shows prediction accuracy equivalent to the subjects’ revisit performance. 

 \end{abstract}


\end{frontmatter}

\section{Introduction}
In many applications of human-robot systems, both humans and robots can execute tasks. 
For some human routine tasks, for example, object pick-up or handling, robots not only have their advantages (e.g. low cost) but disadvantages (e.g. low reliability). 
Humans can easily perfectly do such tasks, but the cost is higher: humans are slower, subject to fatigue, and sparse in the such systems. 
In such applications, a common decision-making problem is ``What is more beneficial, choosing the robot (option) or the human (option)?" 
\cite{mathieu2000influence} proved that when all team members share the same mental model, its overall performance improves. 
Hence, we need to study how humans make decisions and embody the same model on robots since it is neither feasible nor moral to force humans think mechanically.
In particular, human decision-making is inevitably influenced by the regret emotion \citep{zeelenberg2007theory}. 
This influence in decision-making is modeled by regret theory \citep{loomes1982regret}.
Regret theory is a qualitative model which is more for analysis than for prediction as needed for robotic decision-making.
To transform the model from qualitative to quantitative, one requires to reliably measure preference in human minds. 
A well-designed measurement instrument thus is particularly important. 

Measuring the mental values is the common act in psychology and economics. 
Questionnaire-based survey is the predominant tool. 
For questionnaires, designing the instrument is to design the human-computer interface (HCI) if the test is conducted on computers. Hence, these two terms can be used interchangeably. 
Designing a proper HCI is uneasy because our understanding of human perception, heuristics and cognition is limited. 
Any of unintended aspects of the design may influence human judgments in unexpected direction \citep{fischhoff1988knowing}.
Research devoted to survey methodology is abundant \citep{mckelvie1978graphic,lozano2008effect,harpe2015analyze}. 
The main purposes of these works are to avoid ambiguity, refine response, and reduce mental effort. 
The Likert-type scale with 5 points emerges as the widely-accepted optimal one \citep{mckelvie1978graphic,lozano2008effect}.
However, \cite{li2013novel} argued that the standard Likert method has drawbacks such as information loss and distortion. 
She proposed to incorporate fuzzy sets theory to the Likert scale by requiring subjects to report how much degree the rated level matches their opinion. 
Similarly, \cite{fourali1997using} designed a fuzzy instrument that allows subjects to choose more than one scores on a rating scale. 
It relaxes the mental strain of subjects by acknowledging that subjects often know a range other than a single value during rating. 
\cite{de2015fuzzy} proposed a method to require subjects directly draw the membership functions, introduced by fuzzy sets theory, over a range of values.
It gives subjects more freedom in manifesting their thought and feeling. 
These designs all enable more subtle expression. 
However, they unavoidably complicate the process for answering the questions. 

In this work, we contribute a simple, comprehensive, neutral design of the HCI that is compatible with its real world applications. 
The main novelty of this HCI design is, firstly, we utilize the connection between fuzzy sets theory and natural language to measure more detailed data but in an intuitive way to the human subjects; secondly, we consider human heuristics in judgment in an effort to make the measurement less distorted and more reliable. 

The rest of the paper is organized as follows. 
Section \ref{sec:2} formulates the decision-making problem. 
Section \ref{sec:3} explains in detail the design of the HCI. 
Section \ref{sec:4} describes the experiment using the HCI. Section \ref{sec:5} presents
the experimental results. Section \ref{sec:6} concludes the work.

\section{Decision-Making Problem Formulation}
\label{sec:2}
\begin{figure}
\begin{center}
\includegraphics[width=6cm]{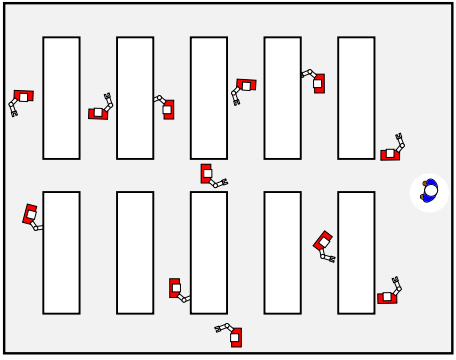}    
\caption{A human robot team in a warehouse (top view).} 
\label{fig:warehouse}
\end{center}
\end{figure}
For the sake of illustration, we choose an application in warehouses without losing generality. 
A human worker acts as an agent collaborating with a group of robots in picking up objects from shelves in a super-market style warehouse as in \figurename\, \ref{fig:warehouse}. 
In normal configuration, the robots are distributed on the floor for picking up while the worker remains in a station. 
The worker has direct communication with every robot such that he/she can respond to robots' requests. 
Robots' grippers are not dexterous enough to guarantee every pick is a success because of the diversity of goods. 
But it is assumed that robots can recognize every good and have a probabilistic estimate of pick-up success rate $p_r$. 
The cost of a successful robot pick-up (the robot option) is negligible.
However, when a pick-up fails, the good drops, causing damage cost $x_r < \$\, 0$. 
On the other hand, the worker can pick up any objects with confident ease---the probabilistic success rate is $1$. 
Such benefit comes with price. 
The worker has to disengage with other work he/she is doing, take time to move to the requesting location, and help the robot. 
Considering human workforce is more expensive, a human pick-up (the human option) requires a non-negligible cost $x_h< \$\, 0$. 
When a robot is ready for a pick-up, it needs to contemplate whether to resort to the robot option, which is risky, or the human option, which imposes a certain moderate cost. 
This decision-making is summarized in \tablename\, \ref{tb:twooption}.
\begin{table}[ht]
\renewcommand{\arraystretch}{1.3}
\caption{Two options faced by a robot}
\label{tb:twooption}
\centering
\begin{tabular}{c c c c c c c}
\multicolumn{3}{c}{\bf  Robot option} & & \multicolumn{3}{c}{\bf  Human option}\\
\cmidrule{1-3} \cmidrule{5-7}
Outcome:   &   $\$\, 0$    &     $x_r$ &  & Outcome:   &  \multicolumn{2}{c} { $x_h$}\\
Probability: & $p_r$ &      $1-p_r$&  & Probability:  & \multicolumn{2}{c} {$100\%$}\\
\cmidrule{1-3} \cmidrule{5-7}
\end{tabular}
\end{table}

The complete application of one-human-multi-robot collaboration involves motion planing, priory assignment, and etc., hence deserves a separate study. 
The decision-making part, however, is identical for each robot. 
We can focus on one robot without loss of generality. 
 
We want the robot to share the same mental model with the worker. 
The robot should know that when facing the same situation how the worker makes decisions. It then acts accordingly. 
Therefore, we model the decision-making of the worker in front of the options in \tablename~\ref{tb:twooption}.  
On the basis of Regret Theory, we propose a human decision-making model as follows, 
\begin{equation}
e_{rh} \triangleq w(p_r)Q( 0-\overline{x}_h  ) + (1 - w(p_r) )Q(\overline{x}_r -\overline{x}_h).
\label{eq:modified_RT}
\end{equation} 
The probability weighting function  $w(\cdot)$ represents the subjective perception of the objective probabilities \citep{gonzalez1999shape}. 
Function $Q(\cdot)$ is an odd function defined in \citep{loomes1982regret}. 
It evaluates the utilities of outcomes and includes the influence of the regret emotion on decision-making. 
(See \citep{loomes1982regret} for details.) 
The normalized cost is defined as $\overline{x}_{(\cdot)} \triangleq x_{(\cdot)}/{|\overline{x}|}$, where $|\overline{x}|$ is the upper bound of the magnitude range for all outcomes, i.e. $|x_{(\cdot)}|\in [0,\, |\overline{x}|]$. 
Variable $e_{rh}$ is the net advantage of the robot option over the human option.
Its sign determines the choice:
\begin{equation}
e_{rh} \, \substack{ >\\ = \\ <}\, 0 \Leftrightarrow r\, \substack{\succ \\ \sim \\ \prec} \, h,
\label{eqn:DecisionMaking}
\end{equation}
where $r\substack{\succ \\ \sim \\ \prec} h$ means that the robot option is preferred to, equally liked as, or surpassed by the human option. 
Moreover, larger magnitude $|e_{rh}|$ indicates stronger preference.

Equation (\ref{eq:modified_RT}) gives only a framework, since the exact $w$-function and $Q$-function are not given. 
Quantitative $w$-function and $Q$-function, however, are required for a complete model. 
To do so, we must use the clue in equations (\ref{eq:modified_RT}) and (\ref{eqn:DecisionMaking}).
The only variable unrelated to the two functions in this equation is $e_{rh}$. 
According to equation (\ref{eqn:DecisionMaking}), $e_{rh}=0$ means that the two options are equally liked. 
Thus, we can eliminate $e_{rh}$ from the equation by studying only the cases when $r \sim  h$.
If we further define new variables $\delta_{i+1} \triangleq  \overline{x}_r-\overline{x}_h$ and $\delta_i \triangleq \overline{x}_h$, equation (\ref{eq:modified_RT}) becomes 
\begin{equation}
Q\left( \delta_{i+1} \right)=\frac{w(p_r^*)}{1-w(p_r^*)}Q\left(\delta_i \right),
\label{eqn:Q_seq_raw}
\end{equation}
where $p_r^*$ is the probability to make $r \sim h$ hold. 

This equation provides a way to iteratively calculate a sequence of points on the $Q$-function when choosing a specific $w$-function from all possible $w$-function candidates (See more details of calculating $Q(\cdot)$ in \citep{liao2017quantitative}). 
For each $w$-function, we can have a $Q$-function. 
Among all the possible pairs of $(w(\cdot),\, Q(\cdot))$, we then can find the optimal one that shows the best fit with human data.
\begin{figure*}[t]
\begin{center}
\includegraphics[width=0.95\textwidth]{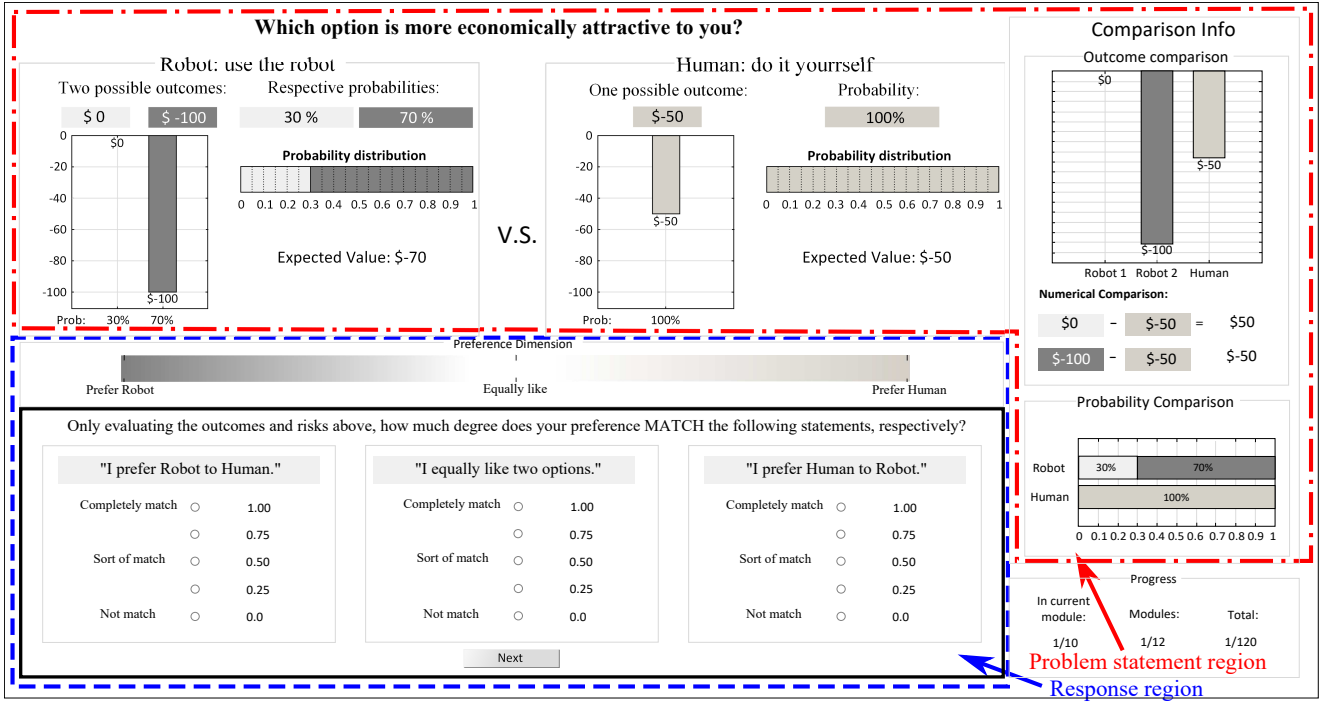} 
\caption{The graphical interface consisting of two regions: the problem statement region enclosed by the red dash-dot line, the response region enclosed by the blue dashed line. (Dashed lines are not shown in the actual HCI.)} 
\label{fig:interface}
\end{center}
\end{figure*}

The prerequisite of the above strategy is $r\sim h$.
Pairs of options coinciding with this preference condition are rare. 
It is then important for a pair of options---no matter what their initial preference condition is---to converge to $r \sim h$. 
It depends on an instrument that can reliably measure the human preference. 
The goal of this work is to design such an instrument (HCI).

\section{Design of the Survey-based HCI}
\label{sec:3}

In designing the HCI, we kept 3 principles in mind \citep{fischhoff1988knowing},
\begin{enumerate}
\item \emph{Compatibility}: Formulate questions in the experiment compatibly with their real-world appearance.
\item \emph{Neutrality}: Avoid implicit biases in information presentation.
\item \emph{Simplicity}: Reduce cognitive workload without destroying subjects' intuition. 
\end{enumerate}
These principles lead to a design of a graphical interface shown in \figurename~\ref{fig:interface}. 
The problem statement region displays the information for a pair of defined options as in \tablename~\ref{tb:twooption}. 
The response region guides subjects in measuring their preferences. 
The interface shows the information of only one decision-making problem.
To elicit the human decision-making model, many such problems are needed. 
The HCI needs a good organization of these problems. 

\subsection{Problem Statement Region}
The problem statement comprises 3 blocks, displaying the question, the individual option information and the option comparison information, respectively. 

The question in the title of \figurename~\ref{fig:interface} asks subjects to directly make a choice. 
It is the same question asked in the warehouse application, avoiding invalid readings due to procedure change \citep{tversky1988contingent}.
The labels of the options---\emph{Robot} and \emph{Human}---have impact on subjects' judgments, because people have different impressions of robots \citep{arras2005we}. 
This influence is necessary because the worker's impression of robots does matter, so we keep the labels, but it should not overshadow the attributes of options---outcomes and probabilities. 
The word \emph{economically} in the title is important because it focuses the attention of subjects on the attributes.

The attributes of one option is nested within one cell (see \figurename~\ref{fig:interface}). 
This type of framing exempts the complaints that the information is distorted by the event-splitting effect \citep{bleichrodt2015regret}. 
To visualize the values of the attributes, both outcomes and probabilities are graphically represented by bar charts. 
The bar chart is chosen for its merit of salient proportionality, making it easier to see the changes of the values. 
Many previous experiments only provides visualization for probabilities (e.g. \citep{bleichrodt2015regret}).
This unbalanced presentation has the potential to introduce bias between different types of attributes. 
To comply with the neutrality principle, it is better to provide visual aids for both attributes. 
Furthermore, the expected value of options is provided, because of two reasons.
Firstly, in pilot tests in which the expected value were not provided, the subjects tried to calculate the expected values mentally. 
It is better to directly provide it to avoid the mental effort.
Secondly, regret theory is originally proposed as an alternative to expected utility theory \citep{loomes1982regret}. 
Since the expected value is the simplest form of expected utility, according to the neutrality principle, it is important to let subjects acknowledge it. 

\cite{loomes1982regret} hypothesized in regret theory that the influence of the regret emotion, which is evoked by the comparison of outcomes in different options, is substantial to decision-making. 
They modeled the comparison as subtraction in equation (\ref{eq:modified_RT}). 
\cite{zadeh1996fuzzy}, however, suggested that it is not natural for human mind to compute numbers. 
The workload for human to do mental calculation is large. 
To reduce it, we provide the comparison information. 
It consists of the comparison of the outcomes and the probabilities. 
Although the probability comparison is not modeled in equation (\ref{eq:modified_RT}), its inclusion exempts potential information bias. 

This is a dilemma between two guiding principles: as discussed above, the comparison information block is prescribed by the principle of simplicity; however, the mere appearance of juxtaposition primes subjects to think in the way biasing toward regret theory \citep{harless1992actions}. 
The compromise is to display the comparison information on the right margin of the interface. 
The right hand side of the page generally attracts little visual attention unless the content there is necessary \citep{buscher2009you}.

\subsection{Response Region}
Preference is the feeling of attractiveness. 
Preference exists objectively, although it cannot be measured as physical signals \citep{tversky1988contingent}. 
\cite{zadeh1996fuzzy} speculated that human minds percept, reason and communicate mainly in natural language rather than in calculation. 
Natural language may be the only proper conveyance. 

\begin{figure}[ht]
\centering
\includegraphics[width = 8cm]{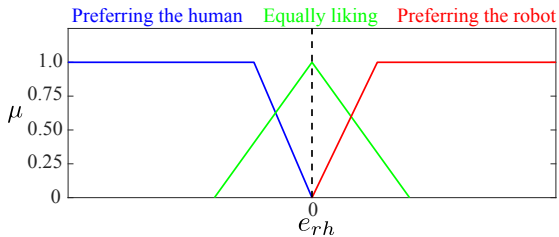}
\caption{The hypothetic definition of fuzzy sets on $e_{rh}$}
\label{fg:HypotheticFuzzy}
\end{figure}

We would like to model the preference expressed in natural language and 
fuzzy sets theory is the tool for this job \citep{zadeh1996fuzzy}. 
A statement in natural language is denoted as a linguistic label and, by definition, is a fuzzy set. 
With respect to the preference, three linguistic labels can be defined: \emph{preferring the robot}, \emph{preferring the human}, and \emph{equally liking}. 
The meaning of the 3 labels is easily understandable. 
People's preference often is more subtle and refined than the 3 categories. 
They often can estimate the degree of match between their inner preference with certain label. 
The degree of match is defined as membership $\mu_l$ to a fuzzy set (linguistic label) $l$. 
Hence, subjects can intuitively and easily describe their preference in a pair $(l,\, \mu_l)$. 
The connection between $(l,\, \mu_l)$ and $e_{rh}$ is the defined membership function $\mu_l(\cdot)$.
Three hypothetic $\mu_l(\cdot)$, one for each linguistic label, are shown in \figurename~\ref{fg:HypotheticFuzzy}.

In the HCI we have to collect $(l, \, \mu_l)$ respectively for the 3 linguistic labels. 
We use an introspection-type of questions to ask subjects to evaluate the degree of match for the statements (see \figurename~\ref{fig:interface}), for example, 
\begin{quote}
\emph{``I prefer Robot to Human.''}
\end{quote}
The degree of match is measured using a 5-point rating scale, since the 5 levels gives the optimal reliability and validity \citep{mckelvie1978graphic} (see the first cell in the response region in \figurename~\ref{fig:interface}). 
We label the scale with both numbers and verbal labels for enhanced scale anchoring \citep{harpe2015analyze}.


The spatial organization of the response region aims to reduce the mental workload. 
The following spatial cues are used: towards left corresponds to preferring the robot; center, equally liking; right, preferring the human; up,  value increasing; down, value decreasing. 
To realize these cues, a gradual colored horizontal bar provides hints to the spatial coordination of preference (See \figurename~\ref{fig:interface}). 
The rating scales querying preferring the robot, equally liking, preferring the human are on the left, center and right, respectively. 
The scales of rating starts from 0 at the bottom to 1 at the top. 


\subsection{Architecture of the HCI}
\figurename~\ref{fig:interface} only represent the information of one decision-making problem and collects their responses. However, many problems of this type should be organized in a sequence to create the architecture of the HCI. 
This architecture determines what problems are included and how they are queued. 
In the designing, we consider relevant aspects of human heuristics---namely, range effect and anchoring effect---to avoid pitfalls due to them. 

\subsubsection{Range Effect.}
Range effect refers to the unwanted phenomenon in experimentation that simply different ranges of attribute values can cause different decision-making \citep{hutchinson1983locus}. 
Outcomes, in our case, do not have a well-defined fixed range, thus are subject to this effect.
For the options in \tablename~\ref{tb:twooption}, the range of outcomes immediately available to subjects is defined by $[x_r,\, 0]$, since $x_r<x_h<0$. 
To combat the range effect, we should maintain a constant range throughout the experiment. 
That to say our objective is to ensure $x_r$ always in a close neighborhood of a constant. 
At the same time, the assignment of values to $x_r$ and $x_h$ should also facilitate the iterative calculation in equation (\ref{eqn:Q_seq_raw}). 
Specifically, our another objective is to find sequences of $x_r$ and $x_h$, respectively, such that $\delta_i$ and $\delta_{i+1}$---both are defined on $x_r$ and $x_h$---construct an extended connected chain $\{\delta_i\}$, e.g. $\{\delta_i\}= \{-0.9,\, -0.8,\,\ldots, \, -0.1\}$. 
One way to achieve both of the above objectives is to define sequences of $\overline{x}_r$ and $\overline{x}_h$ as in \tablename~\ref{tb:Assignment} (recall the definition $\overline{x}_{(\cdot)} \triangleq x_{(\cdot)}/{|\overline{x}|}$).

\begin{table}[ht]
\centering
\caption{The sequences of $\overline{x}_r$ and $\overline{x}_h$}
\label{tb:Assignment}
\begin{tabular}{ccccc}
$i$ &  $\delta_i$ & $\delta_{i+1}$ & $\overline{x}_r$ & $\overline{x}_h$ \\\hline
0 & -0.5 & -0.4 & -0.9 & -0.5 \\
1 & -0.4 & -0.6 & -1.0 & -0.4 \\
2 & -0.6 & -0.3 & -0.9 & -0.6 \\
3 & -0.3 & -0.7 & -1.0 & -0.3 \\
4 & -0.7 & -0.2 & -0.9 & -0.7 \\
5 & -0.2 & -0.8 & -1.0 & -0.2 \\
6 & -0.8 & -0.1 & -0.9 & -0.8 \\
7 & -0.1 & -0.9 & -1.0 & -0.1 \\ \hline
\end{tabular}
\end{table}

Inspecting the columns of $\delta_i$ and $\delta_{i+1}$, there is a connected chain $\{\delta_i\}$ extends from $-0.5$ to $-0.9$, iteratively. 
Also, in the column of $\overline{x}_r$ the value alternates between $-0.9$ and $-1.0$, which is within a close neighborhood of $-1.0$.

\subsubsection{Anchoring Effect.}
The goal of the HCI is to converge a pair of options from any preference state (either $r\succ h$ or $r\prec h$) to $r \sim h$, which makes equation (\ref{eqn:Q_seq_raw}) hold. 
Human responses collected by the HCI, nevertheless, suffer insufficient adjustment from initial states of the option-pair: anchoring effect \citep{tversky1974judgment}. 
The anchoring effect dictates that the free variable $p_r^*$ converged from the state $r\succ h$ is highly like to be different with from the state $r\prec h$. Both directions are subject to insufficient adjustment. 
Averaging the value got from the two directions will cancel out at least partly the insufficiency. 
In the architecture, the converging first starts in one direction, either from $r\succ h$ or $r\prec h$. 
Once the state $r \sim h$ is reached, we get $p_{r,i}^*$.
It then immediately jumps to a new initial state and reaches the state $r \sim h$ again from the opposite direction to get $p_{r,i+1}^*$. 
This iteration can go as many times as possible. 
The final $p_r^*$ is the average. 

\section{Experiment}
\label{sec:4}
We recruited 14 graduate students (3 females) from the Mechanical Engineering Department at Clemson University. 
Among them, 12 datasets, each from one subject, are reported while the other 2 (male) are detected as outliers. 
The incentive to each subject was a \$10 flat payment. 

The experiment lasted around 1.5 hours. 
There were 10 modules and each has 10 decision-making problems. 
Among them 8 modules were for training the model and 2 for validating. 
The validating modules were duplicates and each had 10 different problems. 
One module was inserted at the half of the experiment and the other was at the end. 

To prepare, the experimenter explained the structure of the experiment and basic concepts such as expected value, independent random events. 
The subjects then independently finished a training session, which had only one module. 
Before the testing part, the subjects were surveyed with questionnaire for the value of $|\overline{x}|$. 
Value $|\overline{x}|$ is defined as the amount of money that if lost would cause the subject significantly regret.
Each training module contained problems defined by $\overline{x}_r$ and $\overline{x}_h$ from one row of \tablename~\ref{tb:Assignment}, $|\overline{x}|$ and a free variable $p_r$. 
The problems were presented in the form of \figurename~\ref{fig:interface}. 
If the preference state had not reached $r \sim h$ after 10 problems, the last $p_r$ was taken as the approximation of $p_r^*$.

\section{Result and Discussion}
\label{sec:5}
In this section, we show the performance of the quantitative decision-making model elicited with the designed HCI.
Because building the model relies on human responses collected only through the HCI, we believe that bad design of HCI can hardly generate a satisfactory model. 
In other words, a high-performed model implies an acceptable HCI. 

\begin{figure*}[ht]
\centering
\psfrag{\mu}{$\mu$}
\includegraphics[width=\textwidth]{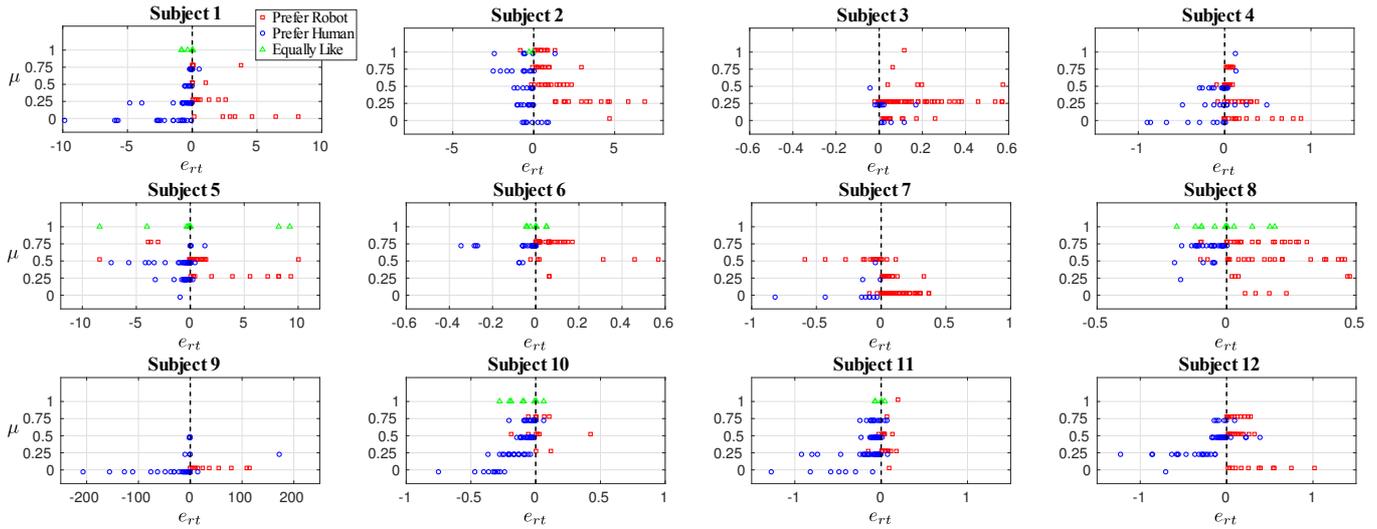}
\caption{The actual membership function of equally liking for individual subject}
\label{fig:AcutalMembership}
\end{figure*}

Since the response region is designed based on fuzzy sets theory, we hypothesized the membership functions in \figurename~\ref{fg:HypotheticFuzzy}.
Using the elicited $Q(\cdot)$ and $w(\cdot)$, we can calculate $e_{rh}$ with equation (\ref{eq:modified_RT}) for any decision-making problem defined by $\overline{x}_r$, $\overline{x}_h$, $|\overline{x}|$ and $p_r$. 
For the same problem, the subjects provided their response $(l, \, \mu_l)$ in the experiment. 
Hence, we can plot the actual degree of membership of each subject for the 80 model training data, as in \figurename~\ref{fig:AcutalMembership}.
For clarity, only the label equally liking---whose membership function in \figurename~\ref{fg:HypotheticFuzzy} is triangular---is considered. 

Regardless of the point type, the shape of the point clouds of 9 subjects resembles a triangle to different degrees (except subject 3, 5, 8). 
It justifies the use of fuzzy sets theory in the HCI design. 
However, the actual membership functions are much noisier. 
Unlike the hypothetical membership functions, $\mu$ in \figurename~\ref{fig:AcutalMembership} spreads over the plotting space. 
It may be due to the fact that human evaluation is probabilistic rather than deterministic. 

According to equation (\ref{eqn:DecisionMaking}), if a point is on the right half plane, the choice should be the robot option; on the left half, the human option; otherwise, equally liking. 
In \figurename~\ref{fig:AcutalMembership}, the points are denoted as indicated in the legend by the subjects' actual response in the experiment. 
The elicited model performs well in the sense that most of the points from the model align with the ground truth. 
However, the plots use the model training data. 
A more objective evidence should come from the validating data.

\begin{figure}[th]
\centering
\includegraphics[width=7.5cm]{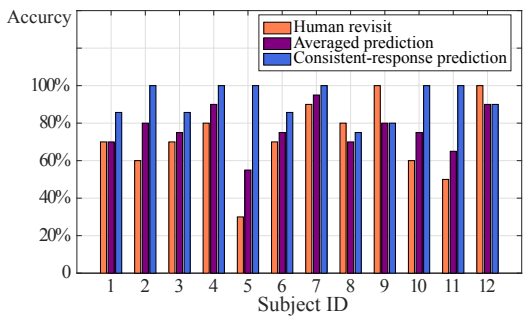}
\caption{Comparison of the model prediction with the subjects' actual choices in validating data}
\label{fig:ModelAccuracy}
\end{figure}

During the experiment, the subjects answered the same validating module twice, their responses in these two modules are compared. 
When two choices to the same decision-making problem are the same, this response is regarded as consistent. 
The percentage of the consistency is named as \emph{human revisit accuracy}. 
The choices predicted by the elicited model are compared with the subjects' actual choices for the two modules, respectively. 
The average percentages of the accurate predictions in one modules are denoted as \emph{averaged prediction accuracy}. 
We further segregate the consistent responses of each subject and denote the percentage of accurate predictions among these consistent responses as \emph{consistent-response prediction accuracy}.
The accuracies are shown for each subject in \figurename~\ref{fig:ModelAccuracy}.

For each subject, the averaged prediction accuracy is close to the human revisit accuracy (maximum difference is 25\%).
The consistent-response prediction accuracy is high ($\geq 75\%$) for each subject. 
This is true even for subject 5 (consistent-response prediction accuracy is 100\%) who has low human revisit accuracy (30\%).
Moreover, the averaged prediction accuracy is strongly positively correlated with the human revisit accuracy (correlation coefficient is 0.79). 
For the whole group, the averaged prediction accuracy ($M=76.67\%$ and $SD=11.35\%$) is close to the human revisit accuracy ($M=71.67\%$ and $SD=20.38\%$). 
A paired-samples t-test shows no significant difference: $t(11)= -1.30$, $p= 0.22$.
The prediction of the quantitative decision-making model can achieve the accuracy as good as the subjects re-answer the problems themselves.  
In terms of consistent-response prediction accuracy, the model performs satisfactory ($M=91.85\%$ and $SD = 9.24\%$). 
The data show that the elicited decision-making model is satisfactory in predicting human decisions, which implies the HCI is also well-performing.

\section{Conclusion}
\label{sec:6}
Including humans in the cyber-physical systems requires modeling human decision-making behaviors. 
The computation units then can either share the model or assist the humans based on the model. 
Such a model needs information from human minds. 
Designing a measurement instrument to do this job is crucial yet not easy because of our limited understanding of human mind. 
But for the known psychological effects of human heuristics, it is important to acknowledge them during designing. 
This work shows that a well-designed HCI can help us elicit a satisfactory computational human decision-making model. 
This elicited model can help improve human-robot collaboration performance  is however still a hypothesis, which will be tested in our future work. 

\bibliography{ifacconf}     

\end{document}